\documentclass[12pt,]{article}
\usepackage{authblk}
\usepackage{natbib}
\usepackage{amssymb,amsmath}
\usepackage[utf8x]{inputenc}
\usepackage{authblk}
\providecommand{\authorinfo}[1]{\noindent{#1}\vspace{2em}}
\providecommand{\keywords}[1]{\noindent\textbf{Keywords: } #1}

\providecommand{\abstractm}[1]{\noindent\textbf{Abstract: } #1\vspace{1em}}
\usepackage{array}
\usepackage{tabularx}
\usepackage{multirow}
\usepackage{quoting}
\usepackage{setspace}
\usepackage[unicode=true]{hyperref}
\usepackage{enumitem}
\usepackage{color}
  \definecolor{mygray}{rgb}{0.9,0.9,0.9}
\usepackage{textcomp}
\usepackage{listings}
\usepackage{hyperref}
\usepackage{graphicx}
\graphicspath{{}}
\usepackage[normal]{subfigure}
\usepackage{array} 
\usepackage{tabularx} 
\usepackage{multirow} 
\usepackage{rotating} 
\usepackage{lscape} 
\title{The \emph{Saltbox-Roof} Probability Distribution}
\author[1]{Ludger O. Suarez-Burgoa}
\affil[1]{Universidad Nacional de Colombia, \href{mailto:losuarezb@unal.edu.co}{losuarezb@unal.edu.co}}
\date{\today}

\begin{document}

\maketitle

\authorinfo{%
Universidad Nacional de Colombia, Faculty of Mines\\
ORCID ID: \href{https://orcid.org/0000-0002-9760-0277}{0000-0002-9760-0277}\\
Present address: Department of Civil Engineering, Cl. 65 \# 78-28, Medell\'in, AN 050034, Colombia\\
Corresponding author (e-mail: \url{losuarezb@unal.edu.co})
}

\abstractm{The saltbox-roof parametric probability distribution is a special case of the triangular distribution, where only one side is truncated. Here it is presented as a single and independent distribution, where the explicit equations are defined for its probability density--, the cumulative distribution--, and the inverse of the cumulative distribution (quantile--) functions as also its random generator. Four parameters are necessary to define it: the lower and the upper limits, the mode, and a shape parameter. Also, the saltbox-roof distribution degenerates into the uniform distribution, into a kind of a trapezoidal distribution and into other special cases of the general triangular distribution, all of them which are related to the domain of the shape parameter within the mode. The mean, median, and the variance are also here expressed by explicit equations. The function equations have been verified with theorems of truncated distributions. Three application examples are exposed.}

\keywords{saltbox-roof, probability distribution, triangular distribution}
\newpage

\section{Introduction}

\begin{figure}[!ht]
  \centering
  \includegraphics[width=0.9\textwidth]{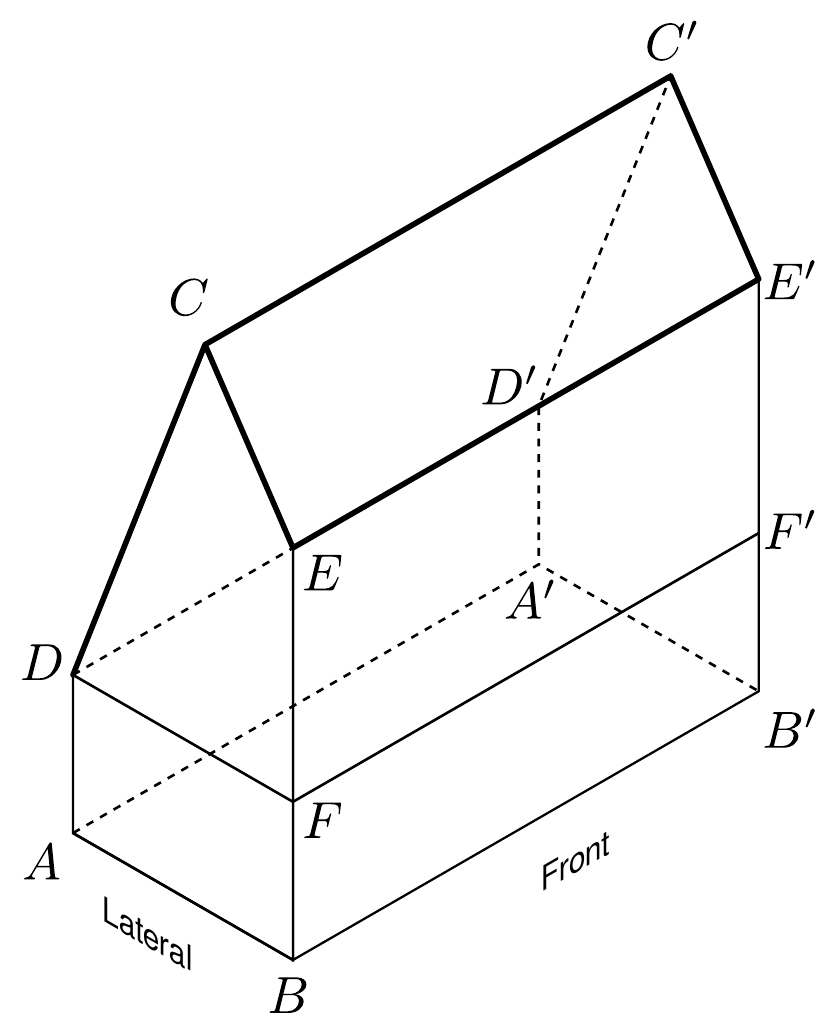}
  \caption{A scheme of a house with a saltbox roof: Plane \(DD'C'C\) and plane \(EE'C'C\).}
  \label{fig:saltboxRoofArchitechture}
\end{figure}

This article deals with a probability distribution that has been called the \emph{Saltbox-Roof} probability density function (PDF), because its shape has a similitude with the roofs that some houses have (Figure \ref{fig:saltboxRoofArchitechture}): In front-view of the house appears to have two stories (plane \(BB'E'E\)) and one short attic (plane \(EE'C'C\)), but in the back-view it has one story (plane \(AA'D'D\)) and a tall attic (plane \(DD'C'C\)). In lateral view, its shape is not symmetric, indeed is a house with two asymmetric hips (lines \(DCE\)) with different slopes.

The here mentioned PDF is not similar to the complete house described above, it is similar to the roof only (without the house's walls); then, the distribution has been named the \emph{Saltbox-Roof} probability distribution.

\begin{figure}[!ht]
  \centering
  \includegraphics[width=1.0\textwidth]{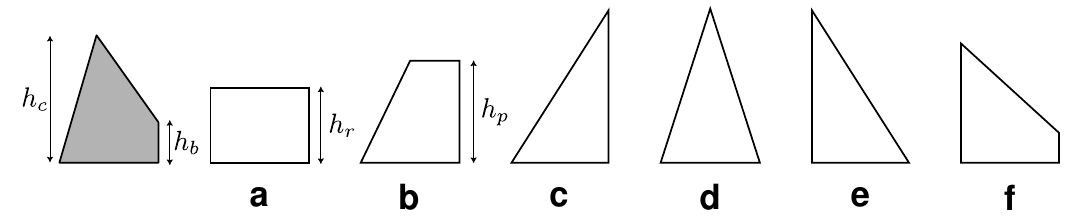}
  \caption{Roof-like probability density functions derived from the Saltbox-Roof PDF (in gray): \textbf{a} Flat roof; \textbf{b} Shed Flat roof; \textbf{c} Left-sided Shed roof; \textbf{d} Gabled roof; \textbf{e} Right-sided Shed roof; \textbf{f} Skillion roof.}
  \label{fig:otherRoofArchitechture}
\end{figure}

This probability distribution degenerates into other simpler PDF whose shapes can also be classified as roofs. The reader will find during this article, that the Saltbox-Roof PDF degenerates into six roof-like PDF, as shown in Figure \ref{fig:otherRoofArchitechture} and described next.

\begin{enumerate}
	\item The Flat-Roof PDF, which is the known \emph{uniform} PDF.
	\item The Gabled-Roof PDF, which is the known \emph{triangular} PDF.
	\item The Right-sided— and Left-sided— Shed-Roofs, which are the triangular PDF in its special case of a right triangle; the first one with its vertical side to the left and the other with its vertical side to the right, respectively. In this case, roof directions are according to the side they drop the water.
	\item The Shed-Flat Roof PDF (also can be called the Shed-Plateau), which possibly it doesn't has an equivalent name in statistics.
	\item The Skillion-Roof PDF (also can be called the Shed-Clerestory), which maybe doesn't has a statistical equivalent name either.
\end{enumerate}

In this article, the explicit equations that defines this Saltbox-Roof distribution is presented: The probability density function, the cumulative density function (CDF), the quantile function (inverse function of the CDF), and the random number generator, among their respective expectation moments such as the mean and the variance.

The validation of the mentioned explicit equations has been made with the property that this Saltbox-Roof distribution is indeed a one \emph{right-sided truncated} triangular distribution; then, theorems of conditional expectations analysis over the triangular distribution has been used for this task.

In the end of this article, three applications examples of the potential use of this distribution is presented.

\section{Equations that defines the Saltbox-Roof distribution}

\begin{figure}[!ht]
  \centering
  \includegraphics[width=0.7\textwidth]{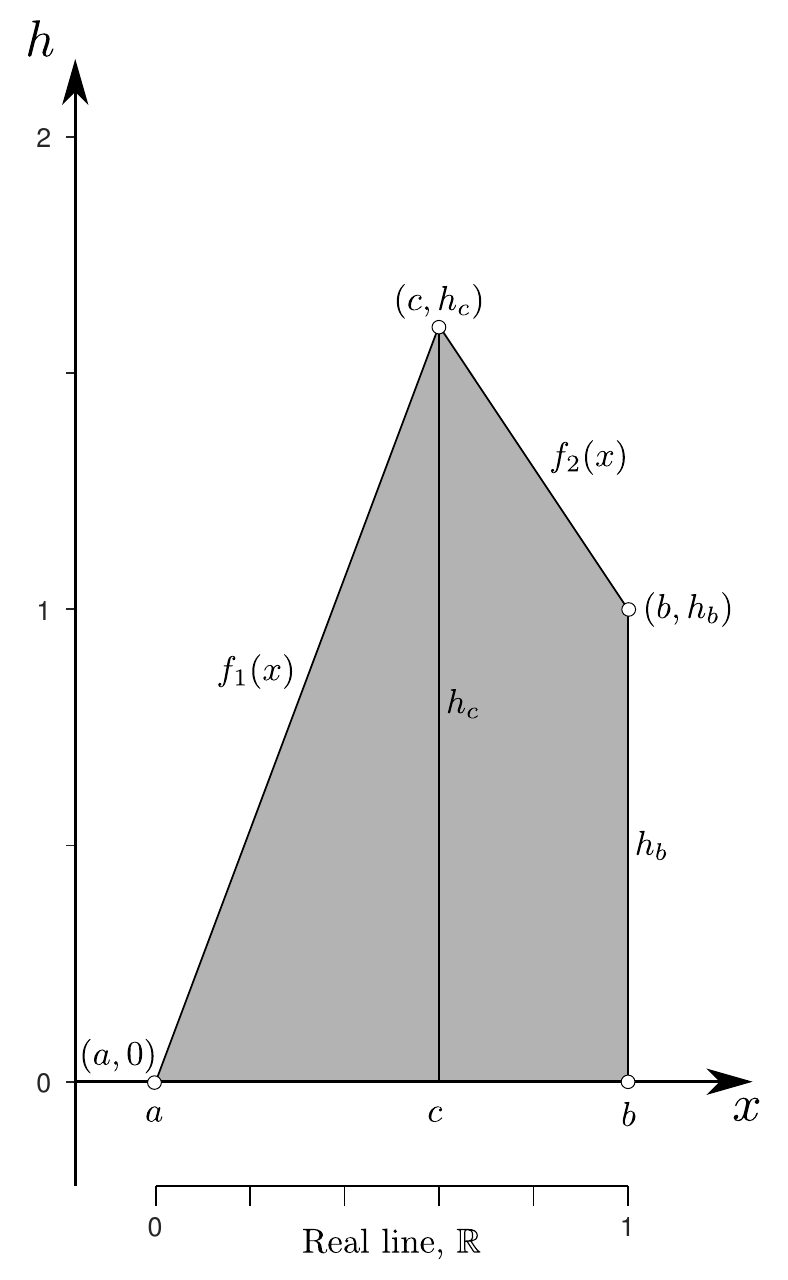}
  \caption{Saltbox-Roof shape as a probability density function, between values \(a\) and \(b\) and mode \(c\).}
  \label{fig:saltboxRoofPDFshape}
\end{figure}

The Saltbox-Roof shape is depicted in a Cartesian two dimensional coordinate system, where the abscissas \(x\) are the quantiles of a \emph{variate value} \(\boldsymbol{X}\) and the ordinates the frequencies \(h = f(x)\), as shown in Figure \ref{fig:saltboxRoofPDFshape}. The variate value is a random variable in the interval of the real line \(\mathbb{R}\).

The equation of the ascending slope, from the left point \(a\) starting from the zero frequency (\(h_a = 0\)) point, \((a, h_a)\) to point \(c\) in the \(h_c\) frequency, \((c, h_c)\), is
\begin{equation}
    \label{eq:f1Line}
    f_1(x) = \frac{x - a}{c - a} h_c.
\end{equation}

The equation of the descending slope, from point \((c, h_c)\) to an end frequency \(h_b\) at \(b\) point, \((b, h_b)\), is
\begin{equation}
\label{eq:f2Line}
    f_2(x) = h_c -\frac{h_c - h_b}{b - a} (x - a).
\end{equation}
The conditions here are that
\[
    h_a = 0,
\]
\[
    h_c \geq h_b,
\]
\[
    a \leq c \leq b.
\]
If those conditions are not accomplished, then the function is not a Saltbox-Roof shape!

\subsection{The probability density function}

To convert the Saltbox-roof shape functions, at Equations \ref{eq:f1Line} and \ref{eq:f2Line}, to a probability density function (PDF), the area under it should be equal to one, \textit{i.e.}
\[
    A_p = \frac{1}{2} (c - a) h_c + \frac{1}{2} (h_c - h_b) (b - c) = 1.
\]

By solving for \(h_b\), it is obtained that
\begin{eqnarray}
    \label{eq:hbFunctionHc}
    \frac{1}{2} (c - a) h_c + \frac{1}{2} (h_c + h_b) (b - c) &=& 1,\nonumber\\ 
    (c - a) h_c + (h_c + h_b) (b - c) &=& 2,\nonumber\\
    h_c [(c - a) + (b - c)] + h_b (b - c) &=& 2,\nonumber\\
    h_b (b - c) &=& 2 - h_c (b - a),\nonumber\\
    h_b &=& \frac{2 - (b - a) h_c}{b - c}.
\end{eqnarray}

By doing some mathematical manipulation in \(f_1(x)\) one gets
\begin{eqnarray*}
    f_1(x) &=& \frac{x - a}{c - a} h_c,\\
           &=& -\frac{a\, h_c}{c - a} + \frac{h_c}{c - a} x.
\end{eqnarray*}

By replacing \(h_b\) in \(f_2(x)\) and performing more mathematical manipulation, one gets
\begin{eqnarray*}
    f_2(x) &=& h_c -\frac{h_c}{b - c} (x - c) + \frac{h_b}{b - c} (x - c), \\
           &=& h_c -\frac{h_c}{b - c} (x - c) + \frac{2 - (b - a)}{(b - c)^2} (x - c), \\
           &=& h_c -\frac{h_c}{b - c} (x - c) + \frac{2 (x - c)}{(b - c)^2} - \frac{b - a}{(b - c)^2} (x - c), \\
           &=& h_c + \frac{2 (x - c)}{(b - c)^2} - \left[\frac{1}{b - c} + \frac{b - a}{(b - c)^2}\right] h_c (x - c), \\
           &=& h_c + \frac{2}{(b - c)^2} x - \frac{2 c}{(b - c)^2} - \frac{h_c}{b - c}(x - c) - \frac{h_c (b - a)}{(b - c)^2} (x - c), \\
           &=& h_c + \frac{2}{(b - c)^2} x - \frac{2 c}{(b - c)^2} - \frac{h_c}{b - c}x + \frac{c\, h_c}{b - c} - \frac{h_c (b - a)}{(b - c)^2} x +\cdots\nonumber\\
           &&\cdots +\frac{h_c (b - a) c}{(b - c)^2}, \\
           &=& \frac{h_c (b - c)^2 - 2c + h_c\,c (b - c) + h_c\,c (b - a)}{(b - c)^2} +\cdots\nonumber\\
           &&\cdots +\frac{2 - h_c (b - a) - h_c (b - c)}{(b - c)^2} x, \\
           &=& \frac{h_c (b - c)^2 - 2\,c + c\,h_c (b - c + b - a)}{(b - c)^2} + \frac{2 - h_c (b - a + b - c)}{(b - c)^2} x, \\
           &=& \frac{h_c (b - c)^2 - 2c + h_c (2b - c - a) c}{(b - c)^2} + \frac{2 - h_c (2b - c - a)}{(b - c)^2} x.
\end{eqnarray*}

Finally, the PDF of the Saltbox-Roof distribution is defined by the following explicit equation:

\begin{eqnarray}
    \label{eq:saltboxroofPDF}
    f(x) &=& 
    \begin{cases}
        -\dfrac{a\, h_c}{c - a} + \dfrac{h_c}{c - a} x, & \text{for $a \leq x \leq c$} \\
        \dfrac{h_c (b - c)^2 - 2\,c + h_c (2b - c - a) c}{(b - c)^2} + \cdots \\ \cdots + \dfrac{2 - h_c (2b - c - a)}{(b - c)^2} x, & \text{for $c < x \leq b$}
    \end{cases}.
\end{eqnarray}

Note that the frequency (\(h_b\)), the left extreme at \(b\) frequency, is not part of the expressions, because it is function of \(h_c\) as shown in Equation \ref{eq:hbFunctionHc}.

\subsection{The cumulative density function}

The cumulative density function (CDF) of the above function is
\begin{eqnarray*}
    F(x) &=& 
    \begin{cases}
        F_1(x) = \int_a^c{f_1(x) \mathrm{d}x}, & \text{for $a \leq x \leq c$}\\
        F_2(x) = \int_c^b{f_2(x) \mathrm{d}x}, & \text{for $c < x \leq b$}
    \end{cases};
\end{eqnarray*}
where by solving the integrals they result in 
\begin{eqnarray*}
    F_1(x) &=& \frac{a^2 h_c - 2 a\, h_c x + h_c x^2}{2 (c - a)},\\
           &=& \frac{a^2 h_c}{2 (c - a)} - \frac{a\, h_c}{c - a} x + \frac{h_c}{2 (c - a)} x^2;
\end{eqnarray*}
\begin{eqnarray*}
    F_2(x) &=&
    \frac{
    \begin{aligned}
    2 c^2 + [x^2(c + a - 2b) h_c + 2]\, + \cdots \\
    \cdots + [2\, c a b - (c + a) b^2] h_c - 2x[\left(c a - b^2\right) h_c + 2 c]
    \end{aligned}
    }{2 (c - b)^2}, \\
           &=& \frac{2\,c^2 + [2 c a b - (c + a) b^2] h_c} {2 (c - b)^2} + \cdots\nonumber\\
           &&\cdots + \frac{[(c + a - 2 b) h_c + 2] x^2 - 2 [(c a - b^2) h_c + 2 c] x} {2 (c - b)^2}, \\
           &=& \frac{2\,c^2 + [2 c a b - (c + a) b^2] h_c} {2 (c - b)^2} -
           \frac{(c a - b^2) h_c + 2 c}{(c - b)^2} x + \cdots\nonumber\\
           &&\cdots + \frac{(c + a - 2 b) h_c + 2}{2 (c - b)^2} x^2.
\end{eqnarray*}

Which in resume is
\label{eq:saltboxroofCDF}
\begin{eqnarray}
    F(x) &=& 
    \begin{cases}
        \dfrac{a^2 h_c}{2 (c - a)} - \dfrac{a\, h_c}{c - a} x + \dfrac{h_c}{2 (c - a)} x^2, & \text{for $a \leq x \leq c$} \\
        \dfrac{2\,c^2 + [2 c a b - (c + a) b^2] h_c} {2 (c - b)^2} - \cdots\\
           \cdots - \dfrac{(c a - b^2) h_c + 2 c} {(c - b)^2} x + \dfrac{(c + a - 2 b) h_c + 2} {2 (c - b)^2} x^2, & \text{for $c \leq x \leq b$}
    \end{cases}.
\end{eqnarray}

\subsection{The inverse of the cumulative density function}

The inverse of CDF, \textit{i.e.} the quantile function, is
\begin{eqnarray*}
    F^{-1}{(U)} &=& 
    \begin{cases}
        F_1^{-1}{(U)}, & \text{for $F(a) \leq U \leq F(c)$}\\
        F_2^{-1}{(U)}, & \text{for $F(c) < U \leq F(b)$}
    \end{cases};
\end{eqnarray*}
where
\begin{eqnarray*}
    F_1^{-1}{(U)} &=& \frac{a\, h_c + \sqrt{2\,U (c - a) h_c}} {h_c}, \\
                  &=& a + \frac{\sqrt{2\,U (c - a) h_c}} {h_c};
\end{eqnarray*}
\begin{eqnarray*}
    F_2^{-1}{(U)} &=& \frac{2c + (c\, a - b^2) h_c - (c - b) \sqrt{
    \begin{aligned}
    (a^2 - 2 a b + b^2) h_c^2 + 4U\, + \cdots\\
    \cdots + 2 h_c [d (U + 1) + c (U - 1) - 2b U]
    \end{aligned}
    }}
    {(c + a - 2\,b) h_c + 2}, \\
                  &=& \frac{(c\, a - b^2) h_c + 2c}{(c + a - 2 b) h_c + 2} - \frac{(c - b)}{(c + a - 2 a) h_c + 2} \cdots\nonumber\\
                  &&\cdots \sqrt{h_c^2\, (a - b)^2  + 2 h_c [a (U + 1) + c (U - 1) - 2 b U] + 4 U}.
\end{eqnarray*}
\begin{eqnarray*}
    F_2^{-1}{(U)} &=& \frac{(c\, a - b^2) h_c + 2c}{(c + a - 2 b) h_c + 2} - \frac{(c - b)}{(c + a - 2 a) h_c + 2} \cdots\\
    &&\cdots \sqrt{h_c^2\, (a - b)^2  + 2 h_c [a (U + 1) + c (U - 1) - 2 b U] + 4 U}.
\end{eqnarray*}

In conclusion
\begin{eqnarray}
    \label{eq:saltboxroofInvCDF}
    F^{-1}{(U)} = 
    \begin{cases}
        a + \dfrac{\sqrt{2U (c - a) h_c}} {h_c}, \hspace{2em}\text{for $F(a) \leq U \leq F(c)$} \\
        \dfrac{(c\, a - b^2) h_c + 2c}{(c + a - 2 b) h_c + 2} -  \dfrac{(c - b)}{(c + a - 2 b) h_c + 2} \cdots\\
        \cdots \sqrt{
        \begin{aligned}
            h_c^2\, (a - b)^2 + \cdots \\ \cdots + 2 h_c \left[a (U + 1) + c (U - 1) - 2 b U\right] + 4 U
        \end{aligned}
        }, \\\text{for $F(c) \leq U \leq F(b)$}
    \end{cases}.
\end{eqnarray}

\subsection{Domain}

The \(x\)-values of the Saltbox-Roof distribution is between its corresponding lower and upper limits defined by \([a, b]\). The mode of the distribution is the \(x\)-value with the highest frequency \(h_c\); thus \(x = c\) is the mode, where \(a \leq c \leq b\). The \emph{residual frequency}, referred here as \(h_b\) and always located at \(x = b\) can't be an arbitrary value, it depends on \(h_c\) and the above x-limits, as described when the probability function was being defined. This residual frequency is ruled by \(h_b = [2 - (b - a) h_c] / (b - c)\), with the condition that \(h_c \geq h_b\).

The frequency at the mode, \(h_c\), can't be an arbitrary value either because its value depends on its location \(c\); then, the existence of a Saltbox-Roof distribution is also related to the paired values \((c, h_c)\). Both values should take any values inside a domain that will be determined here.

To obtain the domain of the pair of variables \((c, h_c)\) it is better to use \emph{relative parameters} of the variables that define the Saltbox-Roof PDF; \textit{i.e.} convert it into a similar PDF shape where the lowest and highest values of \(x\) are transformed to be between \([0, 1]\) and where the frequency at the mode \(h_c\) to be a proportion number (denoted here as \(\rho\), do not confuse the term \emph{proportion} with the population proportion meaning in statistics which is a parameter that describes a percentage value associated with a population) of an interval where \(h_c\) is valid, \textit{i.e.} where \(h_{cm} \leq h_c \leq h_{cM}\), meaning the sub-indexes \(m\) and \(M\) be respectively the minimum and the maximum values of \(h_c\).

The \(x\)-relative value \(x\) with a hat ans expressed as \(\hat{x}\) is 
\[
    \hat{x} = \frac{x - a}{b - a}.
\]
Every variable over a hat will express to be relative to \((b-a)\).
Hence
\[
    \hat{a} = 0,
\]
\[
    \hat{b} = 1,
\]
\[
    \hat{c} = \frac{c - a}{b - a}.
\]

The proportion \(\rho\) according to the above definition is related with the other variables as
\[
    h_c = h_{cm} + \left(h_{cM} - h_{cm}\right) \rho.
\]
When \(\rho = 0\), then \(h_c = h_{cm}\) and when \(\rho = 1\), then \(h_c = h_{cM}\).

The minimum value of \(h_c\) is when the PDF degenerates into an uniform distribution, \textit{i.e.} when \(h_c = h_b\) and \(c \leftarrow [a, b]\). According to the uniform distribution, then
\[
    h_c = h_b = h_r = \frac{1}{b - a},
\]
\[
    h_{cm} = \frac{1}{b - a};
\]

where \(h_r\) is the frequency of a uniform distribution between the same limits \([a, b]\).
It is better to put
\begin{eqnarray*}
    h_{cm} &=& \frac{1}{b - a} d_1,\\
           &=& \frac{d_1}{b - a} 1;
\end{eqnarray*}
where \(d_1\) will be the relative x-interval which is an \emph{unitary x-interval}.

The maximum value of \(h_c\) exist when the PDF degenerates into a triangular distribution, where \(h_c\) can be located at any value of \(c\) between \([a, b]\). It means, that \(h_b = 0\). According to the triangular distribution, it results that
\[
    h_c =  \frac{2}{b - a},
\]
\[
    h_{cM} = \frac{2}{b - a}.
\]
Also, by taken into account the unitary x-interval one obtains
\begin{eqnarray*}
    h_{cM} &=& \frac{2}{b - a} d_1,\\
           &=& \frac{d_1}{b - a} 2.
\end{eqnarray*}

Calling again the definition of \(\rho\),
\begin{eqnarray*}
    h_c &=& h_{cm} + \left(h_{cM} - h_{cm}\right) \rho, \\
        &=& \frac{d_1}{b - a} 1 + \left(\frac{d_1}{b - a} 2 - \frac{d_1}{b - a} 1\right) \rho,\\
        &=& \frac{d_1}{b - a} 1 + \frac{d_1}{b - a} (2 - 1) \, \rho,\\
        &=& \frac{d_1}{b - a} 1 + \frac{d_1}{b - a} 1\, \rho,\\
        &=& 1\, \frac{d_1}{b - a} \left(1 + \rho\right).
\end{eqnarray*}
If the unit \(1\) is named \(h_1\), as being an \emph{unitary frequency}, it results that
\[
   h_c = h_1\, \left[\frac{d_1}{b - a} (1 + \rho)\right];
\]
indeed, clearly a proportion.

Now solving for \(\rho\)
\begin{eqnarray*}
      h_c &=& h_1\, \left[\frac{d_1}{b - a} \left(1 + \rho\right)\right], \\
      \frac{h_c}{h_1} &=& \frac{d_1}{b - a} \left(1 + \rho\right), \\
      \frac{h_c}{h_1} \frac{b - a}{d_1} &=& \left(1 + \rho\right), \\
      \rho &=& \frac{h_c}{h_1} \frac{b - a}{d_1} - 1;
\end{eqnarray*}
also clearly shows a proportion.

Because \(h_1 = 1\) and \(d_1 = 1\) one can express that
\[
    \rho = h_c (b - a) - 1;
\]
with the reminding that their corresponding unitary dimensions are dividing \(h_c\) and \((b - a)\).

Using the relative values of \(h_c\), \(b\) and \(a\), \textit{i.e.} the variable \(\hat{h}_c \), \(\hat{b} \leftarrow 1\) and \(\hat{a} \leftarrow 0\), it results that \((\hat{b} - \hat{a}) = 1\) and that the proportion for the relative values is
\label{eq:pHatFromHcHat}
\begin{eqnarray}
    \hat{\rho} &=& \hat{h}_c (\hat{b} - \hat{a}) - 1,\nonumber\\
            &=& \hat{h}_c - 1;
\end{eqnarray}
being this time \(\hat{\rho}\) as \(\rho\) relative to \((b- a)\) as stated above.

Since relative proportions to \((b- a)\) should be equal in the original and the transformed shapes, then
\[
   \hat{\rho} = \rho,
\]
\[
   \hat{h}_c - 1 = h_c (b - a) - 1,
\]
\[
   \hat{h}_c = h_c (b - a);
\]
where \((b - a)\) can be considered a \emph{scaling factor} to transform from \(h_c\) to \(\hat{h}_c\).

In resume, to transform the parameters of the Saltbox-Roof distribution into its corresponding \emph{x-interval \([0, 1]\)} assume that
\begin{eqnarray*}
        \hat{a} &=& 0,\\
        \hat{b} &=& 1,\\
        \hat{c} &=& \frac{c - a}{b - a},\\
        \hat{h_c} &=& h_c (b - a).
\end{eqnarray*}

Within the Saltbox-Roof distribution expressed in its \([0, 1]\) \(\hat{x}\)-interval, the domain can be expressed in terms of the pair of variables \((\hat{c}, \hat{\rho})\). The variable \(\hat{c}\) may vary within the interval \([0, 1]\), the variable \(\hat{\rho}\) also between the interval \([0, 1]\).

When \(\hat{\rho} = 0\), it means that \(\hat{h}_c\) is the minimum with a value of 1. If in addition \(\hat{c} = 0\), it means that \(\hat{h}_c\) is located at the initial values of \(\hat{x}\). With respect to the shape, the saltbox-roof shape degenerates into a flat-roof shape, \textit{i.e.} the uniform distribution as shown in Figure \ref{fig:otherRoofArchitechture}a.

If \(\hat{c} = 0\) is maintained constant, an increment \(\hat{\rho}\) from 0 to 1, \(\hat{h}_c\) initially being the minimum value of 1, it will grown until its maximum possible value, \textit{i.e.} 2; but to balance that growth, \(\hat{h}_b\) should shrink from 1 until 0; then, the shape starts from being a flat-roof to being a skillion-roof (Figure \ref{fig:otherRoofArchitechture}f) until being a right-sided shed-roof, \textit{i.e.} the right triangular distribution with its maximum at the beginning (Figure \ref{fig:otherRoofArchitechture}e).

When \(\hat{\rho} = 1\), it means that \(\hat{h}_c\) is the maximum with the value of 2 (as mentioned); to be so, \(\hat{h}_b\) should be zero and the shape result into a gabled-roof; this if \(0 \leq \hat{c} \leq 1\). This gabled-roof shape is the triangular distribution (Figure \ref{fig:otherRoofArchitechture}d). By maintaining \(\hat{\rho} = 1\); if \(\hat{c} = 0\), the gabled-roof shape converts into the mentioned right-sided shed-roof shape, \textit{i.e.} the right-triangular distribution with its maximum at the beginning (Figure \ref{fig:otherRoofArchitechture}e); but, if \(\hat{c} = 1\) the gabled-roof shape converts to a left-sided shed-roof shape, \textit{i.e.} the right-triangular distribution with its maximum at the end (Figure \ref{fig:otherRoofArchitechture}c).

In the interval \(0 \leq \hat{c} \leq 1\) but with \(0 \leq \hat{\rho} \leq 1\) one has the saltbox-roof shape, but as mentioned above, there is a limit; that limit is when \(\hat{h}_c = \hat{h}_b\), because one important condition of all the equations defined for this distribution is that \(\hat{h}_c \geq \hat{h}_b\). Therefore, it is of interest to find the location \(\hat{c}_L\) in the interval \(]0, 1[\) as a limit where \(\hat{h}_c = \hat{h}_b\). Under that condition, the saltbox-roof shape degenerates into a shed-flat-roof shape (Figure \ref{fig:otherRoofArchitechture}d). This, with the condition that the areas of both shapes should preserve unitary, maintaining invariant also the values of the locations \(\hat{a}\) and \(\hat{b}\).

To solve this, the area of the saltbox-roof shape \(\hat{A}_{ps}\), is evaluated for \(\hat{h}_b = \hat{h}_c \leftarrow \hat{h}_L\) and \(\hat{c} \leftarrow \hat{c}_L\) as following
\begin{eqnarray*}
    \hat{A}_{ps} &=& \frac{1}{2} (\hat{c} - \hat{d}) \hat{h}_c + \frac{1}{2} (\hat{h}_c + \hat{h}_e) (\hat{e} - \hat{c}), \\
              &=& \frac{1}{2} (\hat{c}_L - \hat{d}) \hat{h}_L + \frac{1}{2} (\hat{h}_L + \hat{h}_L) (\hat{e} - \hat{c}_L), \\
              &=& \frac{1}{2} (\hat{c}_L - \hat{d}) \hat{h}_L + \hat{h}_L (\hat{e} - \hat{c}_L);
\end{eqnarray*}
where \(\hat{h}_L\) is the frequency at the limit and \(\hat{c}_L\) is the position in the x-axis of this frequency.

Because the shape is representing a PDF, \(\hat{A}_{ps} = 1\), then
\[
     1 = \frac{1}{2} (\hat{c}_L - \hat{a}) \hat{h}_L + \hat{h}_L (\hat{b} - \hat{c}_L)
\]
and is solved for \(\hat{h}_L\) with respect to \(\hat{c}_L\), as following
\begin{eqnarray*}
    2 &=& (\hat{c}_L - \hat{a}) \hat{h}_L + 2\hat{h}_L (\hat{b} - \hat{c}_L), \\
    2 &=& \hat{h}_L (\hat{c}_L - \hat{a} + 2 \hat{b} - 2 \hat{c}_L), \\
    \hat{h}_L &=& \frac{2}{2 \hat{b} - \hat{a} - \hat{c}_L}, \\
              &=& \frac{2}{2 \hat{b} - (\hat{a} + \hat{c}_L)}.
\end{eqnarray*}

As being \(\hat{b} = 1\) and \(\hat{a} = 0\), then
\[
    \hat{h}_L = \frac{2}{2 - \hat{c}_L}.
\]

Finally, to convert it into a \(\rho = \hat{\rho}\) value, it is remembered that
\[
    \hat{\rho} = \hat{h}_c - 1
\]
and by analogy
\[
    \hat{\rho} = \hat{h}_L - 1.
\]
Solving for \(\hat{h}_L\) is obtained that
\[
    \hat{h}_L = \hat{\rho} + 1.
\]

Substituting this into the above found expression, it results that
\label{eq:pHatFromP}
\begin{eqnarray}
    \hat{h}_L &=& \frac{2}{2 - \hat{c}_L},\nonumber\\
    \hat{\rho} + 1 &=& \frac{2}{2 - \hat{c}_L},\nonumber\\
    \hat{\rho} &=& \frac{2}{2 - \hat{c}_L} - 1.
\end{eqnarray}

It will be also useful to have the solution for \(\hat{c}_L\), then
\label{eq:cHatLfromPhat}
\begin{eqnarray}
    \hat{\rho} &=& \frac{2}{2 - \hat{c}_L} - 1,\nonumber\\
    (\hat{\rho} + 1) (2 - \hat{c}_L) &=& 2,\nonumber\\
    2 - \hat{c}_L &=& \frac{2}{\hat{\rho} + 1},\nonumber\\
    \hat{c}_L &=& 2 - \frac{2}{\hat{\rho} + 1}.
\end{eqnarray}

The rightest limit of the saltbox-roof distribution is defined by the following inequality
\begin{eqnarray*}
    \hat{c} &\leq& \hat{c}_L, \\ 
            &\leq& 2 - \frac{2}{\hat{\rho} + 1}. 
\end{eqnarray*}
and the lowest limit of the the saltbox-roof distribution is defined by the following inequality
\begin{eqnarray*}
    \hat{\rho} &\leq& \frac{2}{2 - \hat{c}_L} - 1, \\
            &>& 1 - \frac{2}{2 - \hat{c}_L}.
\end{eqnarray*}

The equation that limits the values of \(\hat{c}\) and \(\hat{\rho}\), \textit{i.e.} where the shed-flat-roof distribution exist as in Figure \ref{fig:otherRoofArchitechture}b,  is
\label{eq:pHatClHatFunction}
\begin{eqnarray}
    (\hat{\rho} + 1) (2 - \hat{c}_L) - 2 &=& 0,\nonumber\\
    2 \hat{\rho} - \hat{\rho} \hat{c}_L + 2 - \hat{c}_L - 2 &=& 0,\nonumber\\
    2 \hat{\rho} - \hat{\rho} \hat{c}_L - \hat{c}_L &=& 0.
\end{eqnarray}

\begin{figure}[!ht]
  \centering
  \includegraphics[width=1.0\textwidth]{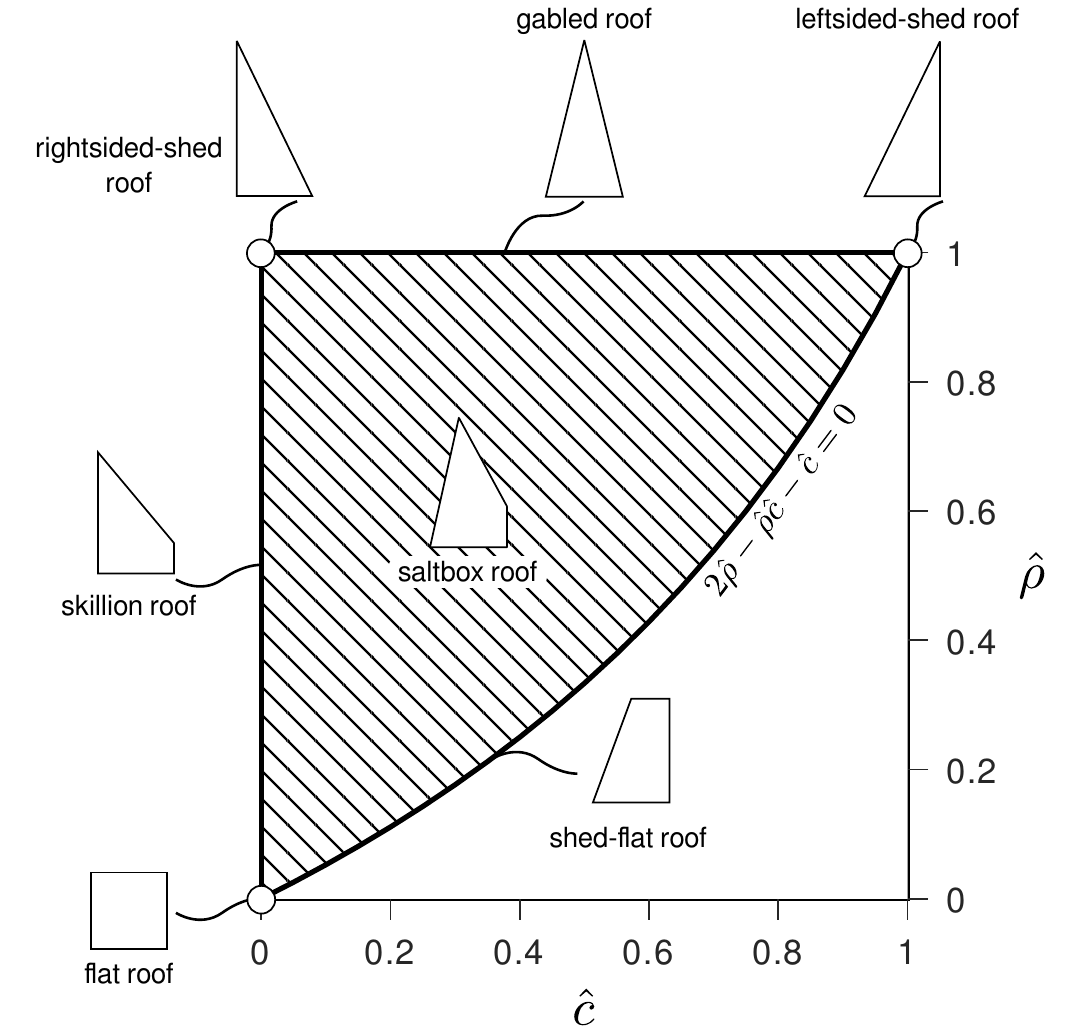}
  \caption{Domain of the Saltbox-Roof distribution.}
  \label{fig:saltboxRoofDomain}
\end{figure}

All the above mentioned limits that define the domain of the existence of the Saltbox-Roof distribution is resumed in Figure \ref{fig:saltboxRoofDomain}.

\subsection{Properties}

In this section it will put the explicit equations of the mean, median, mode, and variance of the Saltbox-roof distribution. These expressions where obtained using a Computer Arithmetic System (CAS), such as SageMath, and were verified numerically. The expressions are the following.

\subsubsection{Mean}
\label{eq:saltboxRoofMean}
\begin{eqnarray}
    \mu &=& -\frac{1}{6} (a^2 - 2 a b + b^2) h_c + \frac{1}{3} c + \frac{2}{3} b,\nonumber\\
        &=& -\frac{1}{6} (a - b)^2 h_c + \frac{1}{3} c + \frac{2}{3} b.
\end{eqnarray}

\subsubsection{Median}
\label{eq:saltboxRoofMedian}
\begin{equation}
    m = \frac{a h_c + \sqrt{(c - a) h_c}}{h_c}.
\end{equation}

\subsubsection{Mode}
\label{eq:saltboxRoofMode}
\begin{equation}
    M = c.
\end{equation}

\subsubsection{Variance}
\label{eq:saltboxVariance}
\begin{eqnarray}
    \sigma^2 &=& -\frac{1}{36} (a^4 - 4 a^3 b + 6 a^2 b^2 - 4 a b^3 + b^4) h_c^2 + \frac{1}{18} c^2 - \frac{1}{9} c b + \frac{1}{18} b^2 + \cdots\nonumber\\
    &&\cdots + \frac{1}{36} (c a^2 - 3 a^3 + (c - 7 a) b^2 + 2 b^3 - 2 (c a - 4 a^2) b) h_c,\nonumber\\
             &=& -\frac{1}{36} (a - b)^4 h_c^2 + \frac{1}{18} c^2 - \frac{1}{9} c b + \frac{1}{18} b^2 + \frac{1}{36} (c - 3a + 2b) (a - b)^2 h_c,\nonumber\\
             &=& -\frac{1}{36} (a - b)^4 h_c^2 + \frac{1}{18} (c - b)^2 + \frac{1}{36} (c - 3a + 2b) (a - b)^2 h_c,\nonumber\\
             &=& \frac{1}{36} \left[2 (c - b)^2 + (c - 3a + 2b) (a - b)^2 h_c - (a - b)^4 h_c^2 \right].
\end{eqnarray}

\section{Distributions that degenerates from the Saltbox-Roof distribution}

As shown in Figure \ref{fig:saltboxRoofDomain} the Saltbox-Roof distribution degenerates to other functions. In this section it is presented the equations of these functions, some of them well known by the mathematical community.

\subsection{Uniform distribution (Flat-Roof)}

The uniform distribution between the interval \([a, b]\) has the same probability for all its values; necessary should have a constant height \(h_r\) to have a rectangular area equal the unity. Then, the rectangle's area is
\[
    A_p = (b - a) h_r = 1.
\]
Solving for \(h_r\) it results that 
\[
    h_r = \frac{1}{b - a}.
\]

The equation of the probability density function in the interval is 
\label{eq:uniformPDF}
\begin{eqnarray}
    f(x) &=& h_r,\nonumber\\
         &=& \dfrac{1}{b - a}, \hspace{2em}\text{for $a \leq x \leq b$}.
\end{eqnarray}

The cumulative density function is
\label{eq:uniformCDF}
\begin{eqnarray}
    F(x) &=& \int_d^x {f(x) \mathrm{d}x},\nonumber\\
         &=& \int_d^x {h_r \mathrm{d}x},\nonumber\\
         &=& \int_d^x {\frac{1}{b - a} \mathrm{d}x},\nonumber\\
         &=& \left. \frac{1}{b - a} x \right\vert _d^x,\nonumber\\
         &=& \frac{x - a}{b - a}, \hspace{2em}\text{for $a \leq x \leq b$}.
\end{eqnarray}

The inverse of the CDF is
\label{eq:uniformInvCDF}
\begin{eqnarray}
    F^{-1}(U) = d + U (b - a);
\end{eqnarray}
where \(U\) is any random number, \textit{i.e.} the same uniform distribution, but between the interval \([0, 1]\).

\subsection{Triangular distribution (Gabled-Roof)}

The PDF function is 
\label{eq:triangularPDF}
\begin{eqnarray}
    f(x) = 
    \begin{cases}
            \dfrac{2(x-a)}{(b-a)(c-a)}, &\text{for \(a\leq x \leq c\)}\\
            \dfrac{2(b-x)}{(b-a)(b-c)}, &\text{for \(c \leq x \leq b\)} 
    \end{cases}.
\end{eqnarray}

To obtain the CDF, the function \(f(x)\) is integrated between the interval \([a, b]\)
\begin{eqnarray*}
    F(x) &=& \int{f(x)} \mathrm{d}x, \\
    &=&
    \begin{cases}
        \int_a^x {\dfrac{2(x-a)}{(b-a)(c-a)}}\mathrm{d}x, &\text{for \(a\leq x \leq c\)}\\
        \int_c^x {\dfrac{2(b-x)}{(b-a)(b-c)}}\mathrm{d}x, &\text{for \(c \leq x \leq b\)} 
    \end{cases}.
\end{eqnarray*}

After the integration is performed, the CDF is
\label{eq:triangularCDF}
\begin{eqnarray}
    F(x) &=& 
    \begin{cases}
        \dfrac{a^2 - 2 \, a x + x^2}{\left(a - b\right) \left(a - c\right)}, &\text{for \(a\leq x \leq c\)}\\
        \dfrac{a^{2} - 2 \, a c + c^{2}}{{\left(a - b\right)} {\left(a - c\right)}} + \dfrac{2 \, b c - c^{2} - 2 \, b x + x^{2}}{{\left(a - b\right)} {\left(b - c\right)}}, &\text{for \(c \leq x \leq b\)} 
    \end{cases}.
\end{eqnarray}

To generate random numbers for this CDF, the inverse of \(F(x)\) is obtained; this by changing \(F(x)\leftarrow U\) and solving for \(x\); \textit{i.e.} \(x\leftarrow F^{-1}\), resulting in the following equations:
\begin{eqnarray*}
   F_1^{-1}(U) &=& a + \sqrt{U a^{2} - U a b - {\left(U a - U b\right)} c}, \\
               &=& a + \sqrt{U} \sqrt{a^2 - a b - (a - b)c}, \\
               &=& a + \sqrt{U} \sqrt{a(a - b) - (a - b)c}, \\
               &=& a + \sqrt{U} \sqrt{(a - b)(a - c)}.
\end{eqnarray*}
\begin{eqnarray*}
   F_2^{-1}(U) &=& b - \sqrt{{\left(U - 1\right)} a b - {\left(U - 1\right)} b^{2} - {\left({\left(U - 1\right)} a - {\left(U - 1\right)} b\right)} c}, \\
               &=& b - \sqrt{U - 1} \sqrt{a b - b^2 - (a - b)c}, \\
               &=& b - \sqrt{U - 1} \sqrt{b (a - b) - (a - b)c}, \\
               &=& b - \sqrt{U - 1} \sqrt{(a - b) (b - c)}, \\
               &=& b - \sqrt{1 - U} \sqrt{(b - a) (b - c)}.
\end{eqnarray*}

Finally, is resumed that 
\label{eq:triangularInvCDF}
\begin{eqnarray}
    F^{-1}(U) = 
    \begin{cases}
        a + \sqrt{U} \sqrt{(b - a)(c - a)}, &\text{for \(F_1(a) \leq U \leq F_1(c)\)} \\
        b - \sqrt{1 - U} \sqrt{(b - a) (b - c)}, &\text{for \(F_2(c) \leq U \leq F_2(b)\)} 
    \end{cases};
\end{eqnarray}
with \(U\) in the interval \([0, 1]\).

\subsection{Right-Triangular distribution}

The right-triangular distribution is a special case of the triangular distribution. We can distinguish two right-triangular distributions, whose maximum frequency is located at the end of the interval (where the slope points to the left) named here the \emph{Left-sided Sheed Roof} distribution and the \emph{Right-sided Shed Roof} distribution, whose slope points to the right and the maximum frequency is located at the beginning of the interval.

\subsubsection{The Left-sided Shed Roof distribution}

The PDF of the Left-sided Shed Roof distribution is found by setting \(c \leftarrow b\) in the equation of the same function for the triangular distribution; then
\label{eq:leftsidedShedRoofPDF}
\begin{eqnarray}
    f(x) = \frac{2(x - a)}{(b - a)^2}, &\text{for $a\leq x \leq b$}.
\end{eqnarray}

The corresponding CDF of this function (using the same rule to set \(c\) equal to \(b\)) is 
\label{eq:leftsidedShedRoofCDF}
\begin{eqnarray}
    F(x) = \frac{(x - a)^2}{(b- a)^2}, &\text{for $a\leq x \leq b$}.
\end{eqnarray}

The inverse of the CDF is therefore 
\label{eq:leftsidedShedRoofInvCDF}
\begin{eqnarray}
    F^{-1}(x) = a + \sqrt{U} (b - a), &\text{for $F(a) \leq U \leq F(b)$}.
\end{eqnarray}

\subsubsection{The Right-sided Shed Roof distribution (Right-sided and Left-sided Shed Roofs)}

Similar to the above case, the PDF of the Right-sided Shed Roof distribution is found by setting \(c \leftarrow a\), then 
\label{eq:rightsidedShedRoofPDF}
\begin{eqnarray}
    f(x) = \frac{2(b - x)}{(b - a)^2}, &\text{for $a\leq x \leq b$}.
\end{eqnarray}

The corresponding CDF of this function (using the same rule to set \(c\) equal to \(a\)) is 
\label{eq:rightsidedShedRoofCDF}
\begin{eqnarray}
    F(x) &=& \dfrac{2ba - a^2 - 2bx + x^2}{(a - b)(b - a)},\nonumber\\
         &=& -\dfrac{2ba - a^2 - 2bx + x^2}{(a - b)^2}, \hspace{2em}\text{for $a\leq x \leq b$}.
\end{eqnarray}

The inverse of the CDF is 
\label{eq:rightsidedShedRoofInvCDF}
\begin{eqnarray}
    F^{-1}(x) &=&  b - \sqrt{(1 - U)(b - a)}, \hspace{2em}\text{for $F(a) \leq U \leq F(b)$}.
\end{eqnarray}

\subsection{One special case of the Trapezoidal distribution (Shed-Flat Roof)}

The functions that defines the \emph{Shed-Flat Roof} distribution are the following
\[
    f_1(x) = \frac{x - a}{c - a} h_p, \hspace{2em}\text{for $a \leq x \leq c$};
\]
\[
    f_2(x) = h_p, \hspace{2em}\text{for $ c \leq x \leq b$}.
\]

The height of the plateau (\(h_p\)) should be determined in function of the variables \({a, b, c}\) to have an unitary area; therefore
\[
    A_p = \frac{1}{2} (c - a) h_p  + (b - a) h_p = 1.
\]
Solving for \(h_p\) is obtained that
\begin{eqnarray*}
    \frac{1}{2} (c - a) h_p  + (b - c) h_p &=& 1, \\
    (c - a) h_p  + 2(b - c) h_p &=& 2, \\
    h_p (c -a + 2b - 2c) &=& 2, \\
    h_p &=& \frac{2}{2b - a - c}, \\
        &=& \frac{2}{2b - (a + c)}.
\end{eqnarray*}

Once replacing \(h_p\) in \(f_1(x)\) and \(f_2(x)\) and performing some math manipulation, one gets
\begin{eqnarray*}
    f_1(x) &=& \frac{x - a}{c - a} \frac{2}{2b - (a + c)},\\
           &=& \frac{2(x - a)}{(c - a)[2b - (c + a)]},\\
           &=& \frac{2(x - a)}{2b(c - a) - (c^2 - a^2)},\\
           &=& \frac{2(x - a)}{(a^2 - c^2) - 2b(a - c)}
\end{eqnarray*}
and
\begin{eqnarray*}
    f_2(x) &=& h_p,\\
           &=& \frac{2}{2b - (a + c)}.
\end{eqnarray*}

The PDF of this distribution is
\label{eq:shedFlatPDF}
\begin{eqnarray}
    f(x) =
    \begin{cases}
        \dfrac{2(x - a)}{(a^2 - c^2) - 2b(a - c)}, &\text{for $a \leq x \leq c$}\\
        \dfrac{2}{2b - (a + c)}, &\text{for $c \leq x \leq b$}
    \end{cases}.
\end{eqnarray}

The CDF is
\label{eq:shedFlatCDF}
\begin{eqnarray}
    F(x) =
    \begin{cases}
        -\dfrac{a^2 - 2\,a x + x^2}{c^2 - a^2 - 2(c - a)e}, &\text{for $a \leq x \leq c$}\\
        -\dfrac{a^2 - 2\,a c + c^2}{c^2 - a^2 - 2(c - a)e} + 2\dfrac{c - x}{c + a - 2b}, &\text{for $c \leq x \leq b$}
    \end{cases}.
\end{eqnarray}

To generate random numbers under this CDF the inverse functions has been determined as following
\label{eq:shedFlatInvCDF}
\begin{eqnarray}
    F^{-1}(x) &=& 
    \begin{cases}
        a + \sqrt{-U c^2 + U a^2 + 2 (Uc - Ua)b}, \\\text{for $F(a) \leq U \leq F(c)$} \\
        -\frac{1}{2} (U - 1)c - \frac{1}{2}(U - 1)a + U b, \\\text{for $F(c) \leq U \leq F(b)$}\\
    \end{cases}.
\end{eqnarray}

\subsection{Another trapezoidal distribution (Skillion Roof)}

In the literature, \emph{e.g.} \cite{Kim2007.inProceedings}, it is defined the trapezoidal distribution as one that has an ascending line up to a limit where it becomes a plateau and then follows a descending line. The present trapezoidal distribution that is dealt here is not that mentioned in the literature.

Here it is defined a slightly different trapezoidal distribution which has a vertical ascending line \(h_c\) at the starting-point of the interval (point \(a\)) and then a descending line up to a vertical line \(h_b\) at the end-point of the interval (point \(b\)); thus \(c = a\). It is a quadrilateral with vertical sides, one side horizontal (the base) and an inclined side opposite to the base (Figure \ref{fig:otherRoofArchitechture}f).

The function that defines the \emph{Skillion Roof} distribution is 
\[
    f(x) = h_c - \frac{h_c - h_b}{b - c} (x - c);
\]
with the condition that
\(h_c \geq h_b\) and \(c \leq b, \) for \(c \leq x \leq b\).

In order to convert it into a PDF, the are should be unitary, then
\[
    A_p = \frac{1}{2} (h_c + h_b) (b - c) = 1.
\]

Solving for \(h_b\) it results that
\begin{eqnarray*}
    \frac{1}{2} (h_c + h_b) (b - c) &=& 1,\\
    h_c (b - c) + h_b (b - c) &=& 2,\\
    h_b &=& \frac{2 - h_c (b - c)}{b - c}.
\end{eqnarray*}

Replacing \(h_b\) into \(f(x)\) one gets
\begin{eqnarray*}
    f(x) &=& h_c - \frac{h_c}{b - c}(x - c) + \frac{h_b}{b - c}(x - c), \\
         &=& h_c - \frac{h_c}{b - c}(x - c) + \frac{2 - h_c (b - c)}{(b - c)^2}(x - c), \\
         &=& h_c + \frac{c\, h_c}{b - c} - \frac{h_c}{b - c} x - \frac{2c - c\,h_c (b - c)}{(b - c)^2} + \frac{2 - h_c (b - c)}{(b - c)^2} x, \\
         &=& \frac{h_c (b - c)^2 + c\,h_c (b - c) + c\,h_c(b - c) - 2c}{(b - c)^2} + \cdots\\
         &&\cdots + \left(\frac{2 - h_c (b - c)}{(b - c)^2} - \frac{h_c}{b - c}\right) x.
\end{eqnarray*}

Finally, the PDF is
\begin{eqnarray*}
    f(x) &=& \frac{h_c (b - c)^2 + 2\,c\,h_c (b - c) - 2c}{(b - c)^2} + \cdots \\
    && \cdots + \left(\frac{2 - h_c (b - c)}{(b - c)^2} - \frac{h_c}{b - c}\right) x, \hspace{2em}\text{for \(c \leq x \leq b\)}.
\end{eqnarray*}

The range of the value of \(h_c\) is maximum when \(h_c \equiv h_b\) turning it into a \emph{Flat-Roof} (uniform) PDF, and minimum when \(h_b = 0\) turning it into a \emph{Right-sided Shed Roof} PDF.

When \(h_c \equiv h_b = h_t\)
\[
    h_c = \frac{1}{b - a}.
\]
When \(h_b = 0\)
\begin{eqnarray*}
    h_b &=& \frac{2 - h_c (b - c)}{b - c}, \\
    0 &=& \frac{2 - h_c (b - c)}{b - c}, \\
    h_c (b - c) &=& 2, \\
    h_c &=& \frac{2}{b - a}.
\end{eqnarray*}

Therefore, those values are a reference for the range of \(h_c\)
\[
    \frac{1}{b - a} \leq h_c \leq \frac{2}{b - a}.
\]

In conclusion, the PDF is
\label{eq:SkillionRoofPDF}
\begin{eqnarray}
    f(x) &=& \frac{h_c (b - a)^2 + 2\, a h_c (b - a) - 2\, a}{(b - a)^2} +\cdots\nonumber\\
    &&\cdots + \left[\frac{2 - h_c (b - a)}{(b - a)^2} - \frac{h_c}{(b - a)}\right] x, \hspace{2em}\text{for $a\leq x \leq b$}.
\end{eqnarray}

Within that PDF, the CDF has been obtained as following 
\label{eq:SkillionRoofCDF}
\begin{eqnarray}
    F(x) &=& \frac{1}{(a - b)^2} \{[(a - b) h_c + 1] x^2 + a^2 + (a^2 b - a b^2) h_c - \cdots\nonumber\\
    &&\cdots - [(a^2 - b^2) h_c + 2\,a] x \}, \hspace{2em}\text{for $a \leq x \leq b$}.
\end{eqnarray}

Finally, the inverse of the CDF of this function is 
\label{eq:SkillionRoofInvCDF}
\begin{eqnarray}
    F^{-1}(U) &=& \frac{1}{2[(a - b) h_c + 1]} \cdots\nonumber\\ 
    &&\cdots \left[2a + (a^2 - b^2) h_c - (a - b) \sqrt{
    \begin{aligned}
    (a^2 - 2\,a b + b^2) h_c^2 + \cdots\\ \cdots + 4\,U (a - b) h_c + 4\,U
    \end{aligned}
    }\right], \nonumber\\
    &&\text{for $F(a) \leq x \leq F(b)$}.
\end{eqnarray}

\section{Validation}

The validation of the explicit equations of the Saltbox-Roof distribution was made by the use of the concept of the \emph{truncated distribution}, used in \emph{conditional probability}, according to \cite{Ushakov2002.inbook} at \cite{Hazenwinkel2002.book}.

\cite{Ushakov2002.inbook} says: Let \(G(x)\) be a cumulative distribution function (the original CDF) between the interval \([d, e]\). The truncated distribution corresponding to \(G(x)\) between \(a\) and \(b\) is understood to be the distribution function
\label{eq:truncatedTriangCDF}
\begin{eqnarray}
    F_{\{a, b\}}(x) = 
    \begin{cases}
        0, & \text{for $x \leq a$}\\
        \dfrac{G(x) - G(a)}{G(b) - G(a)}, &\text{for $a < x \leq b$}\\
        1, & \text{for $x > b$, $a \leq b$}
    \end{cases}.
\end{eqnarray}

Similar holds for the truncated probability density function \(f(x)\)
\label{eq:truncatedTriangPDF}
\begin{eqnarray}
    f_{\{a, b\}}(x) = 
    \begin{cases}
        \dfrac{g(x)}{G(b) - G(a)}, &\text{for $a \leq x \leq b$}\\
        0, &\text{for $x \notin\; ]a, b]$} 
    \end{cases}.
\end{eqnarray}

Also, for the inverse of the CDF it holds that it is equivalent to evaluate \(G^{-1}\) with a variable \(G_a + U [G(b) - G(a)]\), \textit{i.e.}
\label{eq:truncatedTriangInvCDF}
\begin{eqnarray}
    F^{-1}_{\{a, b\}}(U) &=& G^{-1}(U),\nonumber\\ 
        &=& G^{-1}(U \left[G(b) - G(a)\right] + G(a));
\end{eqnarray}
where \(g(x)\) is the probability density function of \(G(x)\) between the interval \([d, e]\).

\begin{figure}[!ht]
  \centering
  \includegraphics[width=0.9\textwidth]{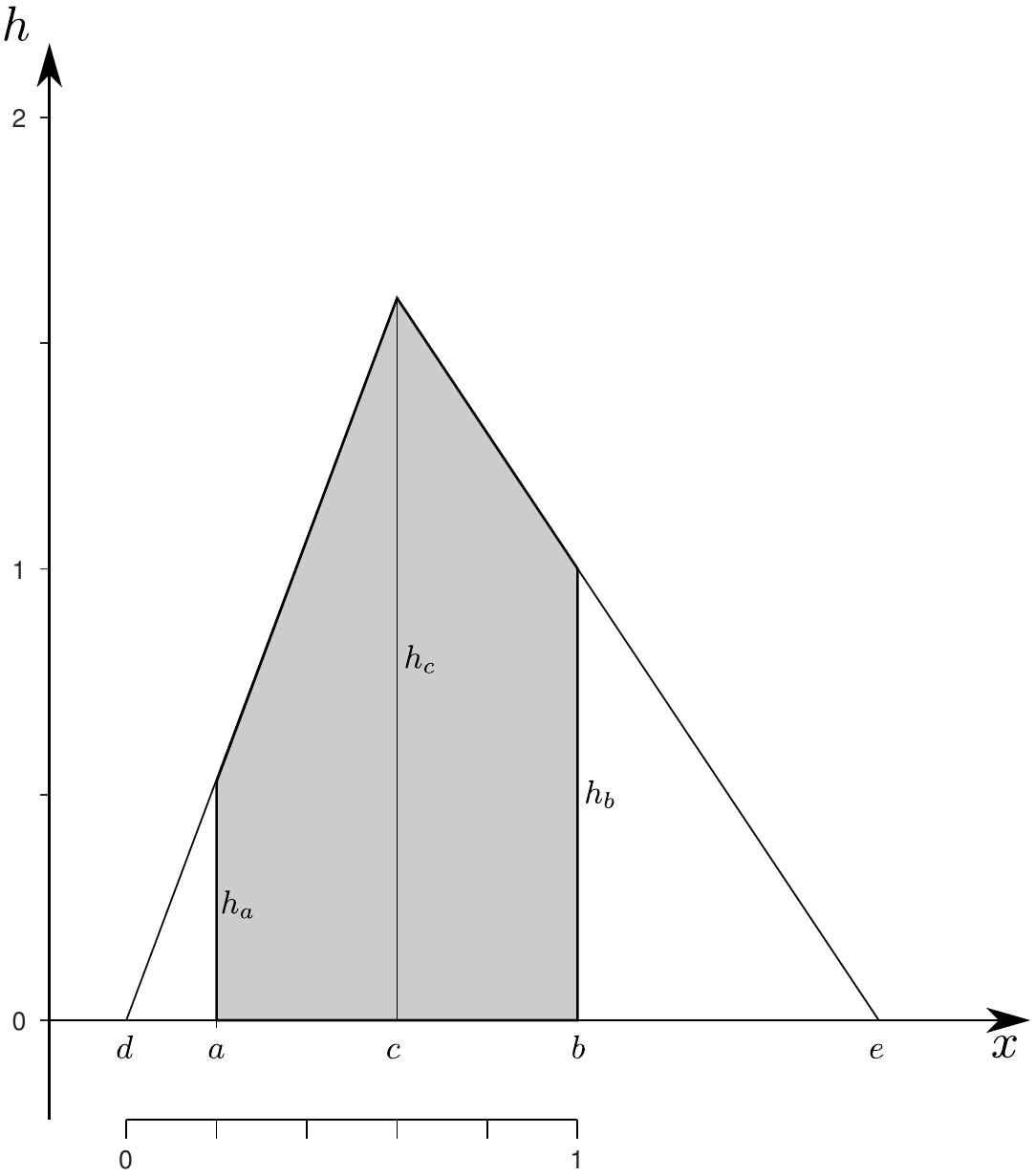}
  \caption{Two sided truncated triangular distribution.}
  \label{fig:truncatedGabledRoofPDFshape}
\end{figure}

In Figure \ref{fig:truncatedGabledRoofPDFshape} is shown this concept graphically for the case of a two-sided truncation. The shape of the original triangle is given by the PDF \(g(x)\), this triangle with base \(\overline{de}\), whose \(G(x)\) and \(G^{-1}(U)\) functions also hold. From this triangle, the \emph{truncated triangle} is the region whose base is \(\overline{ab}\), whose F-functions hold.

\begin{figure}[!ht]
  \centering
  \includegraphics[width=0.9\textwidth]{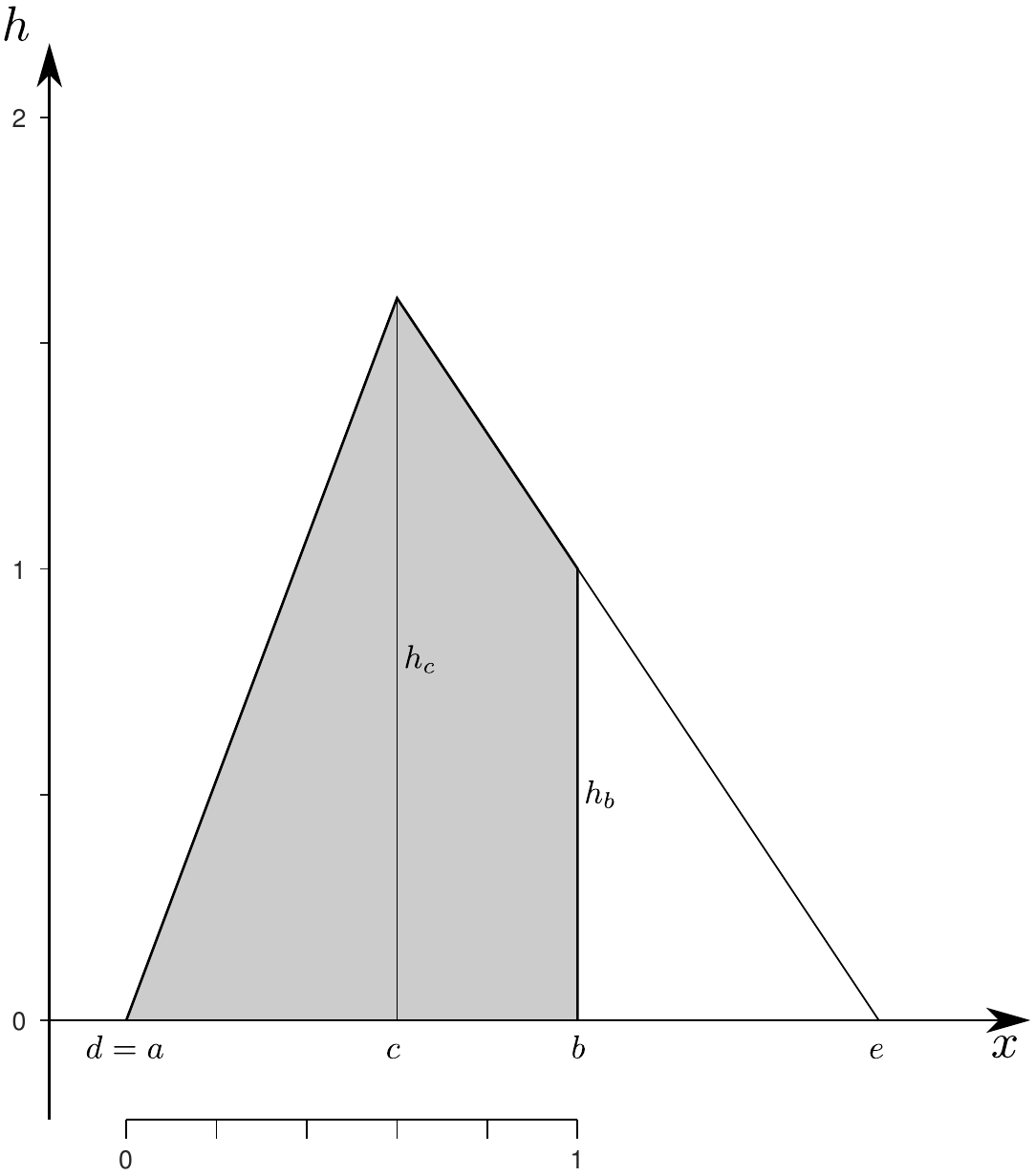}
  \caption{One sided truncated triangular distribution.}
  \label{fig:truncatedOneSideGabledRoofPDFshape}
\end{figure}

If only one side is truncated at \(x \leftarrow b\), such that \(d \leftarrow a\) as shown in Figure \ref{fig:truncatedOneSideGabledRoofPDFshape}; it may say that the triangular distribution is right-sided truncated, \textit{i.e.} the Saltbox-Roof distribution that is dealt in this article.

Similitude between the right-triangles whose heights are \(h_c\) and \(h_b\) is considered to find the value of \(e\) by knowing the values of \(h_b\) and \(h_c\), which result in the following relation
\[
    e = \dfrac{h_c b - h_b c}{h_c - h_b}.
\]

According to this, \(G(a) = 0\) and the above functions are reduced as following:
\begin{eqnarray*}
    f_{\{b\}}(x) = 
    \begin{cases}
        0, &\text{for $x \leq a$}\\
        \dfrac{g(x)}{G(b)}, &\text{for $a \leq x \leq b$}\\
        0, &\text{for $x > b$}
    \end{cases};
\end{eqnarray*}
\begin{eqnarray*}
    F_{\{b\}}(x) = 
    \begin{cases}
        0, &\text{for $x \leq a$}\\
        \dfrac{G(x)}{G(b)} &= \text{for $a \leq x \leq b$}\\
        0, &\text{for $x > b$}
    \end{cases};
\end{eqnarray*}
and
\[
    F^{-1}_{\{b\}}(U) = G^{-1}(U\, G(b)).
\]

For the case of the Saltbox-Roof distribution, \(g(x)\), \(G(x)\) and \(G^{-1}(U)\) are respectively the corresponding PDF, CDF, and the inverse of the CDF; all three of the triangular distribution. In this case, within the interval \([d\leftarrow a, e]\) are respectively:
\label{eq:truncatedSaltboxPDF}
\begin{eqnarray}
    g(x) = 
    \begin{cases}
            \dfrac{2(x-a)}{(e-a)(c-a)}, &\text{for \(a\leq x \leq c\)}\\
            \dfrac{2(e-x)}{(e-a)(e-c)}, &\text{for \(c \leq x \leq e\)} 
    \end{cases};
\end{eqnarray}

\label{eq:truncatedSaltboxCDF}
\begin{eqnarray}
    G(x) = 
    \begin{cases}
        \dfrac{a^{2} - 2 \, a x + x^{2}}{{\left(a - e\right)} {\left(a - c\right)}}, &\text{for \(a\leq x \leq c\)} \\
        \dfrac{a^{2} - 2 \, a c + c^{2}}{{\left(a - e\right)} {\left(a - c\right)}} + \dfrac{2 \, e c - c^{2} - 2 \, e x + x^{2}}{{\left(a - e\right)} {\left(e - c\right)}}, &\text{for \(c \leq x \leq e\)} 
    \end{cases};
\end{eqnarray}
and
\label{eq:truncatedSaltboxInvCDF}
\begin{eqnarray}
    G^{-1}(U) = 
    \begin{cases}
        a + \sqrt{U} \sqrt{(e - a)(c - a)}, &\text{for \(G(a) \leq U \leq G(c)\)} \\
        e - \sqrt{1 - U} \sqrt{(e - a) (e - c)}, &\text{for \(G(c) \leq U \leq G(e)\)} 
    \end{cases}.
\end{eqnarray}

\begin{figure}[!ht]
  \centering
  \includegraphics[width=0.9\textwidth]{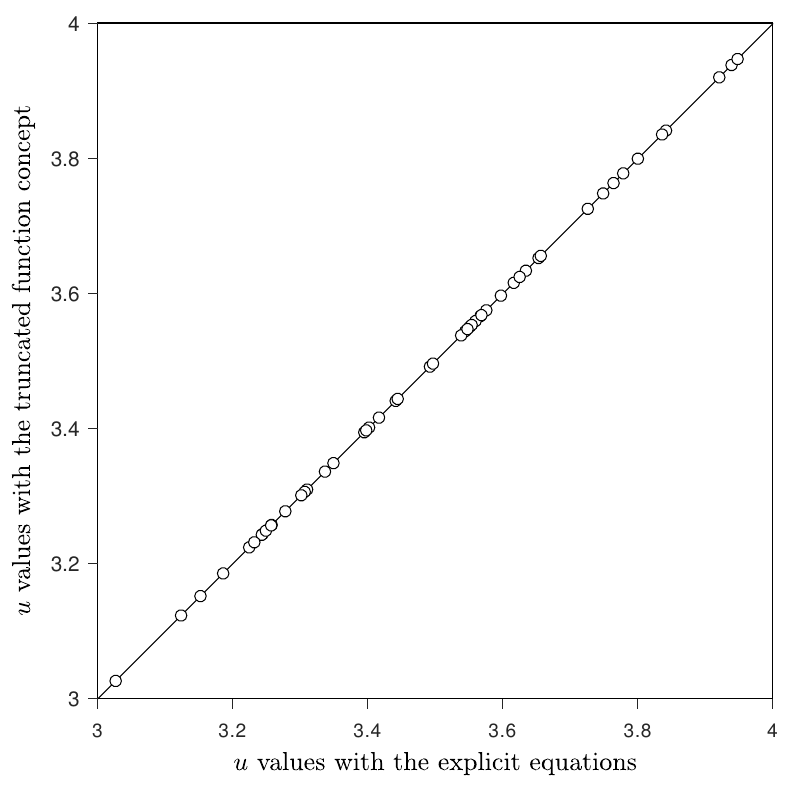}
  \caption{Validation of the Saltbox-Roof distribution.}
  \label{fig:saltboxValidation}
\end{figure}

The validation was made numerically by generating 50 random numbers between \([0, 1]\). With the same random numbers, the explicit function of the inverse CDF of the Saltbox-Roof distribution was used. In Figure \ref{fig:saltboxValidation} is presented the comparison between both approaches. Clearly, they are equivalent.

\section{Applications}

\subsection{Generating random values representing the friction angle of rock surfaces}

The \emph{basic friction angle} (denoted by \(\phi_b\)) between two rock surfaces are those angles obtained by the arc-tangent of the \emph{friction coefficient} at atmospheric Earth pressure, \textit{i.e.} at zero normal-to-the-contact-surface stresses, which is a concept of rock mechanics found for example in \cite{Talobre1957.book}.

Those friction angles can't be negative by physic rules, and they also aren't commonly greater than say around 55 degrees. Indeed, to find rocks with basic friction angles, around 45 degrees, is scarce.

To perform a Monte-Carlo analysis, for example, one can propose a Saltbox-Roof distribution for this basic friction angle, \emph{e.g.} input a minimum of 20 degrees, a maximum of 45 degrees and a mode of 32 degrees. The shape factor \(\hat{\rho}\) could be intuitively preferred be equal to 0.8.

\begin{figure}[!ht]
  \centering
  \includegraphics[width=0.7\textwidth]{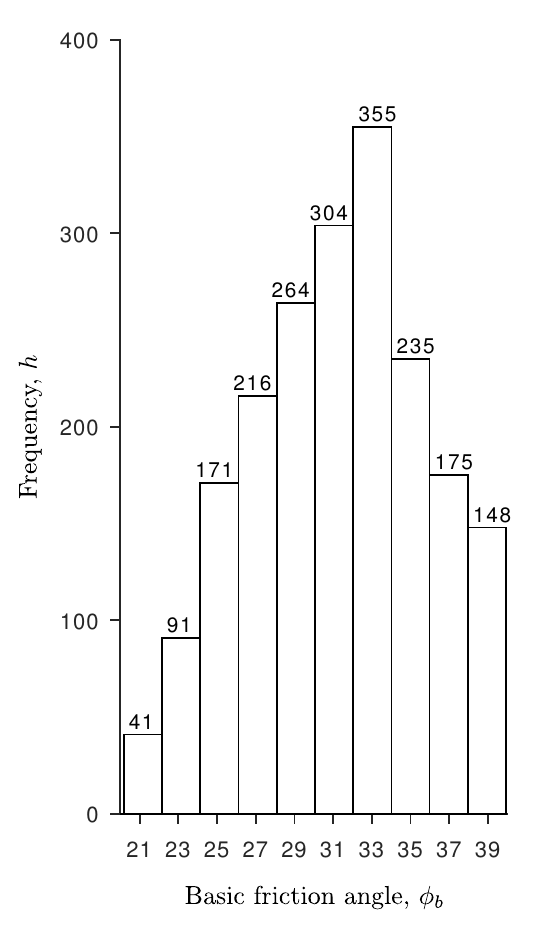}
  \caption{The Saltbox-Roof distribution for the basic friction angle.}
  \label{fig:saltbox4phiB}
\end{figure}

In Figure \ref{fig:saltbox4phiB} is shown the histogram of 2000 generated numbers according to the Saltbox-Roof distribution representing \(\phi_b\).

\subsection{Generating spaced points under a rule in the real-line}

For some computer calculations there is the need to generate equal-spaced points in the set of a one-dimensional space, \textit{i.e.} in \(\mathbb{R}^1\). Many programming languages have functions that generate them; for example, in MATLAB/Octave is available the function named \textsf{linspace} and in Python (in the module Numpy) the function is named the same: \textsf{numpy.linspace}. But among those equal spaced functions there are no more alternatives to control the generation of points in \(\mathbb{R}^1\) apart of the logarithmic version of the linear space, named \textsf{logspace}.

The Saltbox-Roof distribution, and any other probabilistic distribution, can be used to generate \emph{ruled-spaced points} in the real-line; \textit{i.e.} to generate points under an specified rule that is not linear.

Points that are not equal-spaced but not random, they are generated suggested to a rule given by any PDF. In this case, the Saltbox-Roof PDF was be used for that purpose.

\begin{figure}[!ht]
  \centering
  \includegraphics[width=0.9\textwidth]{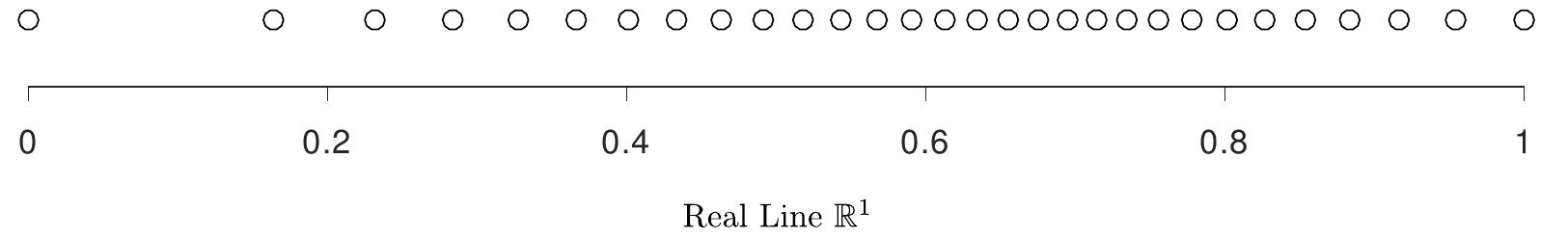}
  \caption{Points spaced by rule of the Saltbox-Roof distribution.}
  \label{fig:saltbox4spacedPoints}
\end{figure}

In Figure \ref{fig:saltbox4spacedPoints} is shown the generation of 30 spaced points in \(\mathbb{R}^1\) in the interval \([0, 1]\) according to the Saltbox-Roof distribution with mode at 0.7 and shape factor of 0.8.

To avoid the randomness of the generated points, it is important to give equal spaced points in the interval \([0, 1]\) and pass them through the inverse CDF function of the PDF; in this case, they are passed through the function in Equation \ref{eq:saltboxroofInvCDF}.

\subsection{Approximating plane curves by polygons with unequal sides}

Plotting curves in the plane (in \(\mathbb{R}^2\)) with variable curvature during its evolution are approximated in principle by polygons, specially in vectorial plotting. The normal method to plot curves is: Generate an equal spaced sequence of numbers that represent the independent variable or the parameter of the curve's function; then, pass those values through the curve's function to get the values of the dependent variable; finally, plot the curves in a 2D Cartesian coordinates system with the values of the independent variable in abscissas (horizontal rightwards x-axis) and the dependent variable in ordinates (vertical upwards y-axis).

In the intervals of the curve where its \emph{curvatures} are small, the resulted polygons are representing the curve smoothly; but in the intervals where the curvatures are high, the curve is shown as polygons rather than a smooth curve.

A common practice to solve this is to increment the number of equal spaced points to obtain smoother curves in the intervals of high curvatures, but in the intervals of low curvatures, those points are unnecessary too much. Then, a non-equal-spaced point generator will save a lot of points in the parts of the curve that has small curvatures and it will concentrate points in the places where the curve has high curvatures.

One accurate algorithm may consider to concentrate points in a direct relation with the different curvatures the curve has, but that will imply to calculate the function of the curve's curvature previous to the generation of the same curve. Sometimes, the curvature functions are cumbersome to obtain. A faster but not an accurate solution may be the use of a flexible distribution generator of points that can save many unnecessary points to define the curve. This can be approximated for example with the inverse function of the CDF of the Saltbox-Roof distribution (or other distribution's inverse CDF).

For this application, consider the vertical parabola \(y = x^2\) with vertex at origin, with parameters \(a_2 = 1\); \(a_1 = a_0 = 0\) for a general parabola \(y = a_2 x^2 + a_1 x + a_0\). The signed curvature \(r^{-1}\) is 
\[
    r^{-1}(x) = \frac{2 a_2}{\left[1 + (2 a_2 x + a_1)^2\right]^\frac{3}{2}};
\]
where by substituting its values \(\{a_0, a_1, a_2\}\) gives the equation of the curvature function with respect \(x\),
\begin{eqnarray*}
    r^{-1}(x) &=& \frac{2}{\left[1 + (2x)^2\right]^\frac{3}{2}}, \\
              &=& \frac{2}{\left(1 + 4x^2\right)^\frac{3}{2}}.
\end{eqnarray*}

\begin{figure}[!ht]
  \centering
  \subfigure[]{\label{subfig:verticalParabolaCurve} \includegraphics[width=0.45\textwidth]{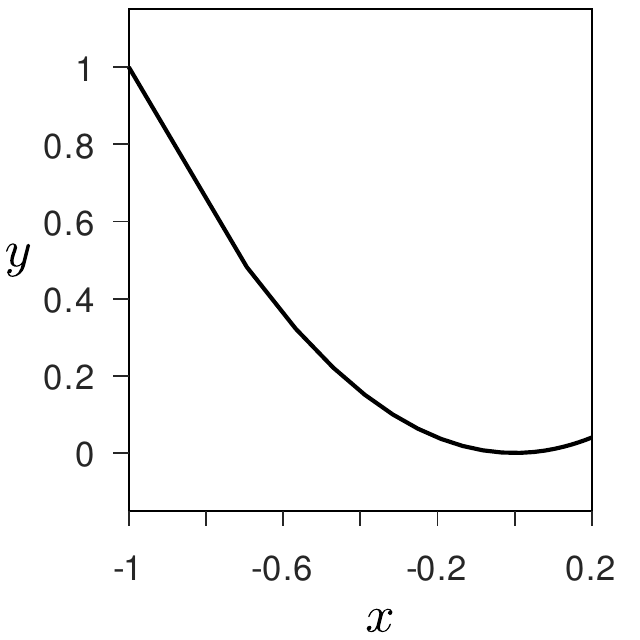}}\quad
  \subfigure[]{\label{subfig:verticalParabolaCurvature} \includegraphics[width=0.45\textwidth]{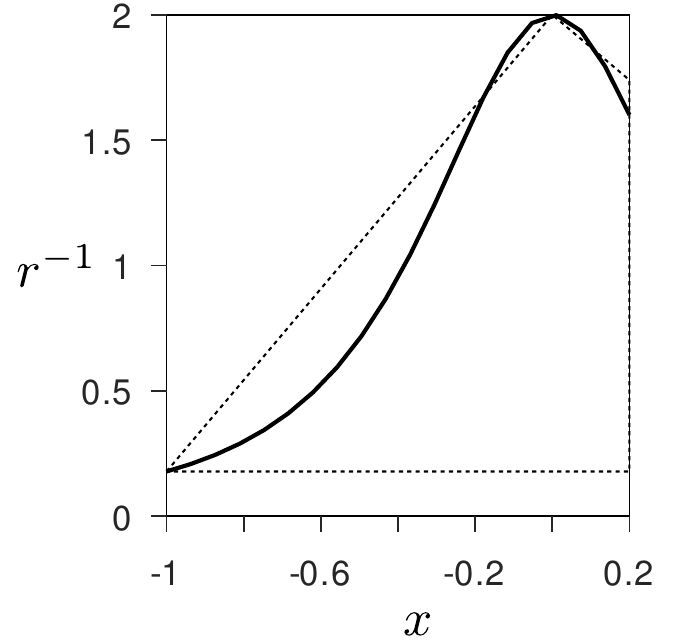}}\\
  \caption{The vertical parabola with vertex at origin \(y = x^2\) in the interval \([-1, \frac{1}{5}]\): \textbf{a} the curve; \textbf{b} the curvature \(\frac{1}{r}\) and with dot line the saltbox-roof distribution shape.}
  \label{fig:verticalParabolaCurveCurvature}
\end{figure}

If the parabola is wanted to plot between the interval \([-1, \frac{1}{5}]\) as shown in Figure \ref{fig:verticalParabolaCurveCurvature}a, one will note that it has a curvature that veries in this interval, as shown in the parabola's curvature \(\frac{1}{r}\) at Figure \ref{fig:verticalParabolaCurveCurvature}b.

Using the plot of Figure \ref{fig:saltboxRoofDomain} about the domain of the Saltbox-Roof distribution, choose the relative pair of parameters \((\hat{c}, \hat{\rho})\) that will give a similar shape as in the parabola's curvature plot, as shown in doted-line in Figure \ref{fig:verticalParabolaCurveCurvature}b.

For example, consider to use the relative mode location at \(\frac{5}{6}\), \textit{i.e.} \(\hat{c} = \frac{5}{6}\); and the shape factor \(\hat{\rho} = \frac{3}{4}\). The choose of \(\hat{c}\) is because the maximum curvature in the plot has an abscissa of \(x=0\), which is located at \(\frac{5}{6}\) the interval width of \(\frac{6}{5}\) length. The choice of \(\hat{\rho}\) is more intuitive, but it is considered that as \(\hat{\rho}\) approaches to zero, the \emph{mode} (at the highest frequency \(h_c\)) and the \emph{residual mode} (at the residual frequency \(h_b\)) are close to be equal.

With that parameters and using the Saltbox-Roof distribution inverse CDF, 20 points are generated and stored in \(\hat{U}\). To transform those values from the interval \([0, 1]\) into the interval \([-1, \frac{1}{5}]\) is used the following equation
\[
    U = x_m + (x_M - x_m) \hat{U};
\]
where \(x_m\) and \(x_M\) are the minimum and maximum values in the interval \([-1, \frac{1}{5}]\), respectively; \textit{i.e.} \(x_m \leftarrow -1\) and \(x_m \leftarrow \frac{1}{5}\).

\begin{figure}[!ht]
  \centering
  \subfigure[]{\label{subfig:saltbox4parabolicCurve} \includegraphics[width=0.45\textwidth]{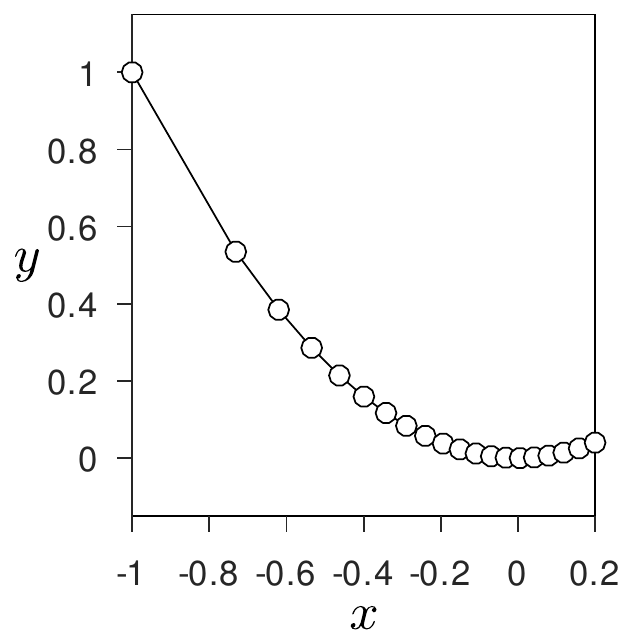}}\quad
  \subfigure[]{\label{subfig:uniform4parabolicCurve} \includegraphics[width=0.45\textwidth]{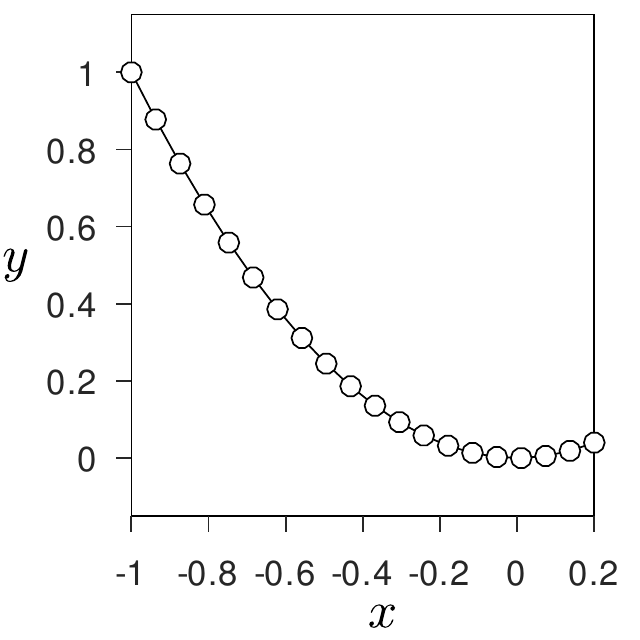}}\\
  \caption{The vertical parabola curve approximated by a 20 points polygon: \textbf{a} points generated with the Saltbox-Roof distribution; \textbf{b} points generated with the Uniform distribution (Flat-Roof).}
  \label{fig:genPts4parabolicCurve}
\end{figure}

Those \(U\) values are passed through function \(y = U^2\), as being \(x \leftarrow U\). The resulting plot is shown in Figure \ref{fig:genPts4parabolicCurve}a. Note that points are concentrated near the parabola's vertex, where the curvature is high. In Figure \ref{fig:genPts4parabolicCurve}b is shown the same parabola in the same interval as plotted  with the same number of points by generating x-values only with the uniform distribution.

\section{Closure}

The Saltbox-roof distribution and its corresponding degenerated cases presented in this document define a set which are intimately related according to which paired of values \((\hat{c}, \hat{\rho})\) one can choose for any case. Figure \ref{fig:saltboxRoofDomain} is a graphical  representation of these relations, where the domain described there hold for any possible saltbox-roof distribution. With this graph and the equations of the probability functions presented here, the user can have control of the shape changes the PDF can have in this set.

\section{Abbreviations}

\begin{description}
   \item[PDF] Probability Density Function.
   \item[CDF] Cumulative Density Function.
   \item[CAS] Computer Arithmetic System.
   \item[2D] Two Dimensions.
\end{description}

\section{Symbols}

\begin{description}
   \item[\(a\)], Inferior limit (minimum value) of \(x\) of the Saltbox-roof distribution.
   \item[\(\hat{a}\)], Inferior relative limit of \(x\).
   \item[\(\{a_0, a_1, a_2\}\)], Set of polynomial coefficients of a 2\(^{nd}\) order polynom \(y(x) \).
   \item[\(A_p\)], Area of a PDF.
   \item[\(\hat{A}_{ps}\)], Are of a PDF in relative variables when \(\hat{h}_c = \hat{h}_b\).
   \item[\(b\)], Superior limit (maximum value) of \(x\) of the Saltbox-roof distribution.
   \item[\(\hat{b}\)], Superior relative limit of \(x\).
   \item[\(c\)], Mode at \(x\) of the Saltbox-roof distribution.
   \item[\(\hat{c}_L\)], Location of \(\hat{c}\) where \(\hat{h}_c = \hat{h}_b\).
   \item[\(\hat{c}\)], Relative mode of \(c\).
   \item[\(d_1\)], Unitary length of an interval.
   \item[\(d\)], Inferior limit of \(x\) in \(g(x)\), \(G(x)\), or \(G^{-1}(x)\).
   \item[\(\overline{de}\)], Length between \(d\) and \(e\).
   \item[\(e\)], Superior limit of \(x\) in \(g(x)\), \(G(x)\), or \(G^{-1}(x)\).
   \item[\(f_{\{a,b\}}(x)\)], PDF of a truncated function between the limits \([a,b]\).
   \item[\(f_{\{b\}}(x)\)], PDF of a right one-sided truncated function between \([a,b]\).
   \item[\(f(x)\)], Function of a PDF.
   \item[\(F_{\{a,b\}}(x)\)], CDF of a truncated function between the limits \([a,b]\).
   \item[\(F_{\{b\}}(x)\)], CDF of a right one-sided truncated function between \([a,b]\).
   \item[\(F(x)\)], Function of a CDF.
   \item[\(F^{-1}_{\{b\}}(x)\)], Inverse CDF of a right one-sided truncated function between  \([a,b]\).
   \item[\(F^{-1}(x)\)], Function of an inverse CDF.
   \item[\(F^{-1}_{\{a,b\}}(x)\)], Inverse CDF of a truncated function between the limits \([a,b]\).
   \item[\(g(a)\)], \(g(x)\) evaluated in \(a\).
   \item[\(g(b)\)], \(g(x)\) evaluated in \(b\).
   \item[\(g(x)\)], PDF of an non-truncated function evaluated in \(x\).
   \item[\(G(a)\)], \(G(x)\) evaluated in \(a\).
   \item[\(G(b)\)], \(G(x)\) evaluated in \(b\).
   \item[\(G(x)\)], CDF of an non-truncated function evaluated in \(x\).
   \item[\(G^{-1}(a)\)], \(G^{-1}(x)\) evaluated at \(a\).
   \item[\(G^{-1}(b)\)], \(G^{-1}(x)\) evaluated at \(b\).
   \item[\(G^{-1}(x)\)], Inverse function of \(G(x)\) evaluated at \(x\).
   \item[\(h\)], Frequency.
   \item[\(h_1\)], Unitary value of frequency.
   \item[\(h_a\)], Frequency at \(a\).
   \item[\(h_b\)], Frequency at \(b\), the residual frequency of the Saltbox-roof distribution.
   \item[\(h_c\)], Frequency at \(c\), the highest frequency of the Saltbox-roof distribution.
   \item[\(\hat{h}_c\)], Frequency at the relative value \(\hat{c}\).
   \item[\(h_{cM}\)], Maximum possible value of \(h_c\).
   \item[\(h_{cm}\)], Minimum possible value of \(h_c\).
   \item[\(\hat{h}_L\)], Frequency at \(\hat{c}_L\).
   \item[\(h_p\)], Plateau frequency in an Shed-flat roof distribution.
   \item[\(h_r\)], Frequency constant in an Uniform distribution.
   \item[\(r^{-1}\)], Signed curvature of \(y\).
   \item[\(\mathbb{R}^{2}\)], Set of the Cartesian product of the set \(\mathbb{R}\).
   \item[\(\mathbb{R}\) or \(\mathbb{R}^{1}\)], Set of real numbers.
   \item[\(m\)], Median.
   \item[\(u\)], Uniform random variable between \([0, 1]\).
   \item[\(U\)], Uniform random function generator between \([0, 1]\).
   \item[\(x\)], Quantil.
   \item[\(\hat{x}\)], Relative quantil.
   \item[\(x_m\)], Minimum value of \(x\).
   \item[\(x_M\)], Maximum value of \(x\).
   \item[\(\boldsymbol{X}\)], Variate value.
   \item[\(y\)], Planar curve function, \emph{i.e.} \(y(x)\) defined in canonical form.
   \item[\(\mu\)], Mean.
   \item[\(\phi_b\)], Basic friction angle (a rock-mechanics concept).
   \item[\(\rho\)], Shape factor of the Saltbox-roof distribution.
   \item[\(\hat{\rho}\)], Relative shape factor.
   \item[\(\sigma^2\)], Variance.
\end{description}

\renewcommand\refname{References}
\bibliographystyle{chicagoa}
\bibliography{references.bib}

\end{document}